\documentclass[aps,pra,12pt]{revtex4-2}
\usepackage[english]{babel}
\usepackage{graphicx,dcolumn,bm,amssymb,amsmath,amsfonts,amsthm,latexsym}
\usepackage{epstopdf}
\usepackage{ulem}
\usepackage{float,subfig,slashed,xcolor,hyperref}
\usepackage[toc]{appendix}
\hypersetup{
  linktocpage  = true,
  colorlinks   = true,
  urlcolor     = red,
  linkcolor    = black,
  citecolor    = blue
}

\begin{document}

\title{Cold isospin asymmetric baryonic rich matter in nonlocal NJL-like models}
\author{J.P.~Carlomagno$^a$}
\email{carlomagno@fisica.unlp.edu.ar}
\author{D.~G\'omez~Dumm$^a$}
\author{N.N.~Scoccola$^{b,c}$}
\affiliation{$^{a}$ IFLP, CONICET $-$ Departamento de F\'{\i}sica, Facultad de Ciencias Exactas,
Universidad Nacional de La Plata, C.C. 67, 1900 La Plata, Argentina}
\affiliation{$^{b}$ CONICET, Rivadavia 1917, 1033 Buenos Aires, Argentina}
\affiliation{$^{c}$ Physics Department, Comisi\'{o}n Nacional de Energ\'{\i}a At\'{o}mica,
Avenida del Libertador 8250, 1429 Buenos Aires, Argentina}

\begin{abstract}
We study the features of low energy strong interactions for a system at zero
temperature and finite baryon and isospin chemical potentials, in the
framework of a Nambu--Jona-Lasinio-like model that includes nonlocal
four-point interactions. We analyze the phase transitions corresponding to
chiral symmetry restoration and pion condensation, comparing our results
with those obtained from local NJL-like models and lattice QCD calculations.
\end{abstract}

\maketitle

\section{Introduction}
\label{intro}

Over the past few decades, there has been a significant amount of research
focused on the study of quark and hadronic matter under conditions of finite
temperature $T$ and baryon chemical potential $\mu_B$. At high temperatures
and low densities, it is well known that quantum chromodynamics (QCD)
predicts the formation of a quark-gluon plasma
(QGP)~\cite{Fukushima:2010bq}, in which quark and gluons are expected to be
weakly coupled. In this limit, strong interactions can be described through
perturbative calculations based on expansions in powers of the QCD coupling
constant. In the region of intermediate temperatures one can rely on lattice
QCD (LQCD) calculations, which indicate that at vanishing chemical potential
the transition from the hadronic phase to the QGP occurs in the form of a
smooth crossover~\cite{Borsanyi:2020fev}. On the other corner of the
$\mu_B-T$ phase diagram, at sufficiently high densities and low
temperatures, one expects to find a ``color-flavor locked'' phase, in which
the existence of strongly correlated quark pairs is
predicted~\cite{Alford:2007xm}. At moderate densities, however, the
situation is much more uncertain. The main reason for this is that
first-principle nonperturbative QCD calculations at nonzero $\mu_B$ are
hardly accessible by Monte Carlo simulations, due to the presence of a
complex fermion determinant in the corresponding partition function (the
so-called ``sign problem'')~\cite{Karsch:2001cy}. In this region most
theoretical analyses of the phase structure rely on the predictions from
effective models for strong interactions.

In addition to $T$ and $\mu_B$, the system may show an imbalance in the
isospin charge, which can be characterized by an isospin chemical potential
$\mu_I$. This situation, which can be applicable e.g.~to the study of the
physics of heavy ion collisions and the structure of stellar objects, is
worth to be considered in order to get more insight into the properties of
strongly interacting matter. In general, the QCD phase diagram in the
$T-\mu_B-\mu_I$ thermodynamic space is expected to show a rich structure
that can be addressed both from LQCD techniques and effective approaches to
strong interactions~\cite{Son:2000xc, Splittorff:2000mm, Loewe:2002tw, Toublan:2004ks, Andersen:2006ys, Kamikado:2012bt, Xia:2013caa, Stiele:2013pma, Ueda:2013sia, Cohen:2015soa, Andersen:2015eoa, Zhang:2015baa, Klahn:2016uce, Kashiwa:2017yvy, Wang:2017vis, Brandt:2017oyy,Brandt:2022hwy,Aryal:2020ocm}.
In the case of systems at $\mu_B=0$ and finite
$\mu_I$, LQCD calculations are not affected by the sign
problem~\cite{Alford:1998sd}; thus, the corresponding phase diagram in the
$T-\mu_I$ plane has been studied in several works that use different lattice
techniques~\cite{Kogut:2004zg,deForcrand:2007uz,Cea:2012ev,Detmold:2012wc,Brandt:2017oyy, Abbott:2023coj}.
In particular, one important feature confirmed by
these calculations is that at $\mu_I \simeq m_\pi$ one finds the onset of a
Bose-Einstein pion condensation phase~\cite{Son:2000xc,Mannarelli:2019hgn},
which could enable the existence of pion stars~\cite{Brandt:2018bwq}.

In the case of nonzero $\mu_B$ and $\mu_I$, lattice analyses are not free
from the sign problem and require some extrapolations. Hence, it is
remarkably important to get definite predictions from effective models. In
this work we study the properties of quark matter under finite $\mu_B$ and
$\mu_I$ conditions considering quark models in which the fermions interact
through covariant {\it nonlocal} four-point couplings~\cite{Ripka:1997zb}.
These models models can be viewed as improved versions of the standard
(local) Nambu$-$Jona-Lasinio (NJL)
scheme~\cite{Nambu:1961fr,Nambu:1961tp}. In fact, nonlocal interactions
naturally emerge within most approaches to low energy QCD, leading to a
momentum dependence in quark propagators that can be successfully reconciled
with LQCD results~\cite{Noguera:2008cm}. Moreover, these so-called
``nonlocal NJL'' (nlNJL) models do not exhibit some of the drawbacks
observed in the local approach. For example, through the usage of
well-behaved nonlocal form factors it is possible to regularize ultraviolet
loop integrals while preserving anomalies~\cite{RuizArriola:1998zi} and
ensuring proper charge quantization. In addition, the absence of sharp
cutoffs implies that model predictions are more stable against changes in
the input parameters~\cite{Blaschke:1995gr}. Within this framework, in a
previous work~\cite{Carlomagno:2021gcy} we have studied the phase diagram in
the $\mu_I-T$ plane for $\mu_B=0$, finding a good agreement with LQCD
calculations. In the present article our aim is to extend this research
considering a system at zero temperature and finite $\mu_I$ and $\mu_B$. We
study the condensate formation and the corresponding phase transitions,
comparing our findings with those obtained from the local NJL
model~\cite{Andersen:2007qv,Liu:2021gsi,Liu:2023uxm,Liu:2016yid}.

In addition, within the nlNJL model we analyze the behavior of the speed of
sound $c_s$ as a function of the isospin chemical potential. Recent LQCD
calculations~\cite{Abbott:2023coj} have found that $c_s^2$ reaches a maximum
at intermediate values of $\mu_I$ ($\mu_I \sim 2m_\pi$), and then decreases
slowly towards the limit predicted by duality for 4D conformal field
theories~\cite{Cherman:2009tw}. This result has been discussed in the
framework of several effective models, see
Refs.~\cite{Chiba:2023ftg,Cao:2020byn,Kojo:2021hqh,Mu:2010zz,Ayala:2023mms}.

This article is organized as follows. In Sec.~\ref{model} we present the
general formalism to describe two-flavor nlNJL models at zero temperature
and finite baryon and isospin chemical potential, including theoretical
expressions for chiral and pion condensates and for the speed of sound. In
Sec.~\ref{results} we discuss our numerical results for condensates and
phase transitions, for both vanishing and nonvanishing $\mu_B$. Finally, in
Sec.~\ref{summary} we summarize our results and present our main
conclusions.

\section{Theoretical Formalism}
\label{model}

We consider a two-flavor quark model that includes nonlocal scalar and
pseudoscalar quark-antiquark currents. The Euclidean action
reads~\cite{Carlomagno:2021gcy}
\begin{eqnarray}
S_E &=& \int d^4 x  \,\left[
    \bar \psi (x) \left(
    - i \rlap/\partial + \hat m  \right) \psi (x)
    \, - \, \frac{G}{2}\, j_a(x) j_a(x) \right] \ ,
\label{action}
\end{eqnarray}
where $\psi = (\psi_u\ \psi_d)^T$ stands for the $u$, $d$ quark field
doublet, and $\hat m = \mbox{diag}(m_u,m_d)$ is the current quark mass
matrix. For simplicity, we assume that the current quark masses $m_u$ and
$m_d$ are equal and we denote them generically by $m_c$. The nonlocal
currents $j_a(x)$ in Eq.~(\ref{action}) are given by
\begin{eqnarray}
j_a (x) &=& \int d^4 z \  {\cal G}(z) \ \bar \psi\left(x+\frac{z}{2}\right) \
\Gamma_a \ \psi\left(x-\frac{z}{2}\right) \ , \label{cuOGE}
\end{eqnarray}
where we have defined $\Gamma_a = ( \openone, i \gamma_5 \vec \tau )$,
$\tau_i$ being Pauli matrices that act on flavor space. The function ${\cal
G}(z)$ is a form factor responsible for the nonlocal character of the
four-point interactions. The action for the standard (local) two-flavor
quark version of the NJL model is recovered by taking ${\cal G}(z) =
\delta^{(4)}(z)$.

To study strongly interacting matter in a system at nonzero chemical
potential we introduce the partition function $\mathcal{Z} = \int
\mathcal{D} \bar{\psi}\,\mathcal{D}\psi \,\exp[-S_E]$. As stated, we are
interested in studying isospin asymmetric matter; this can be effectively
implemented by introducing quark chemical potentials $\mu_u$ and $\mu_d$
that in general are different from each other. Thus, we consider the
effective action in Eq.~(\ref{action}) and perform the replacement
\begin{equation}
\left(\begin{array}{cc} \partial_4 & 0 \\ 0 & \partial_4
\end{array} \right) \ \rightarrow \ \left(\begin{array}{cc} \partial_4 -
\mu_u & 0 \\ 0 & \partial_4 - \mu_d \end{array} \right) \ .
\label{kinrep}
\end{equation}
In fact, it is convenient to write the quark chemical potentials in terms of
average and isospin chemical potentials denoted by $\mu$ ($=\mu_B/3$) and
$\mu_I$, respectively. One has
\begin{equation}
\mu_u \ = \ \mu + \dfrac{\mu_I}{2} \ ,\qquad\qquad \mu_d \ = \ \mu -
\dfrac{\mu_I}{2}\ .
\end{equation}
In addition, owing to the nonlocal character of the interactions, to obtain
the appropriate conserved currents one has to complement the replacement in
Eq.~(\ref{kinrep}) with a modification of the nonlocal currents in
Eq.~(\ref{cuOGE}). This procedure is similar to the one used e.g.\ in
Refs.~\cite{GomezDumm:2006vz,Dumm:2010hh,Carlomagno:2021gcy}.

We proceed now by carrying out a standard bosonization of the effective
theory, introducing bosonic degrees of freedom $\sigma$ and $\pi_i$,
$i=1,2,3$, and integrating out the fermionic fields. Then we consider a mean
field approximation (MFA) in which the bosonic fields are replaced by their
vacuum expectation values (VEVs) $\bar\sigma$ and $\bar\pi_i$. As is well
known, in the chiral limit ($m_c=0$), for $\mu_I=0$ the action is invariant
under global ${\rm U(1)}_B \otimes {\rm SU(2)}_I \otimes {\rm SU(2)}_{IA}$
transformations. The group U(1)$_B$ is associated to baryon number
conservation, while the chiral group SU(2)$_I \otimes {\rm SU(2)}_{IA}$
corresponds to the symmetries under isospin and axial-isospin
transformations. Now, in the presence of a nonzero isospin chemical
potential, the full symmetry group is explicitly broken down to the
U(1)$_{I_3} \otimes {\rm U(1)}_{I_3A}$ subgroup. If $\sigma$ develops a
nonzero VEV, the U(1)$_{I_3A}$ symmetry gets spontaneously broken. Moreover,
while even for finite current quark masses one has $\bar
\pi_3=0$~\cite{Ebert:2006uh}, for $\mu_I\neq 0$ it can happen that $\pi_1$
and $\pi_2$ develop nonvanishing VEVs, leading to a spontaneous breakdown of
the remaining U(1)$_{I_3}$ symmetry. Since the action is still invariant
under U(1)$_{I_3}$ transformations, without loss of generality one can
choose $\bar\pi_i=\delta_{i1}\bar\Delta$.

We consider the above described general situation in which both $\bar\sigma$
and $\bar\Delta$ can be nonvanishing. The mean field grand canonical
thermodynamic potential is found to be given by
\begin{eqnarray}
\Omega^{\rm MFA} &=& \frac{\bar \sigma^2 + \bar\Delta^2}{2\, G}
- {\rm Tr} \ln
\begin{pmatrix}
\rlap/ p_u + M\big(p_u\big) & i \, \gamma_5 \, \rho\big(\bar p\big) \\
i\, \gamma_5 \, \rho\big(\bar p\big)  & \rlap/ p_d + M\big(p_d\big)
\end{pmatrix} \ ,
\label{actionMF}
\end{eqnarray}
where
\begin{eqnarray}
M\big(p\big) \ = \ m_c + g\big(p\big)\, \bar\sigma \ ,  \qquad \qquad
\rho\big(p\big) \ = \ g\big(p\big) \, \bar\Delta \ .
\label{defsMrhop}
\end{eqnarray}
Here we have defined $p_{f\nu} \equiv \left( \vec p ,\, p_4 + i \mu_f
\right)$, with $f=u,d$, and $\bar p = (p_u+p_d)/2$. The function $g(p)$ is
the Fourier transform of the form factor ${\cal G}(z)$ in Eq.~(\ref{cuOGE}).

As usual in this type of model, it is seen that $ \Omega^{\rm MFA}$ turns
out to be divergent and has to be regularized. We adopt here a prescription
similar as the one considered e.g.\ in
Refs.~\cite{GomezDumm:2004sr,Carlomagno:2021gcy}, in which one subtracts the
thermodynamic potential obtained for $\bar\sigma = \bar\Delta =0$ and adds
it in a regularized form. In this way, the regularized expression
$\Omega^{\rm MFA, reg}$ is given by
\begin{equation}
\Omega^{\rm MFA, reg} \ = \
\frac{\bar\sigma^2 + \bar\Delta^2}{2\ G}  - N_f N_c
\int \frac{d^4 p}{(2\pi)^4}\ \ln\bigg[\frac{D(\sigma,\Delta)}{D(0,0)}
\bigg]\, + \, \Omega^{\rm free,reg}\ ,
\end{equation}
where
\begin{equation}
D(\sigma,\Delta) \ = \ E_{u}^2\ E_{d}^2 -\ \rho(\bar p)^2
\Big[\big(M(p_{u})-M(p_{d})\big)^2-(\mu_u-\mu_d)^2\Big] \ ,
\label{granp0}
\end{equation}
with $E_{f}^2=M(p_{f})^2 + p_{f}^2+\rho(\bar p)^2$. The regularized form of
the free piece, after subtraction of divergent terms, reads
\begin{eqnarray}
\Omega^{\rm free,reg} \ = \
-\frac{N_c}{\pi^2} \sum_{f,s=\pm 1} \int_0^\infty p^2 dp\ |\epsilon_f + s\ \mu_f|\ \Theta(-\epsilon_f - s\ \mu_f) \ ,
\end{eqnarray}
where $\epsilon_f = \sqrt{\vec p^{\;2} + m_f^2}\,$.

The mean field values $\bar \sigma$ and $\bar \Delta$ can now be obtained from a set of two coupled
``gap equations'' that follow from the minimization of the regularized
thermodynamic potential, namely
\begin{equation}
\frac{\partial \Omega^{\rm MFA,reg}}{\partial \bar \sigma} \ = \ 0\ , \qquad
\frac{\partial \Omega^{\rm MFA,reg}}{\partial \bar \Delta} \ = \ 0\ .
\label{gapeqs}
\end{equation}

Quark-antiquark and pion condensates are also relevant quantities, since
they can be taken as order parameters of the spontaneous symmetry breaking
transitions. As usual, we consider the scalar condensate $\Sigma = \Sigma_u
+ \Sigma_d$, where $\Sigma_f = \langle \bar \psi_f \psi_f \rangle$; the
latter can be obtained by differentiating the thermodynamic potential with
respect to the current up and down current quark masses, namely
\begin{equation}
\Sigma_f \ = \ \frac{\partial \Omega^{\rm MFA,reg}}{\partial m_f} \ .
\label{Sigma}
\end{equation}
For $\mu_I \neq 0$ one can also have nonvanishing pseudoscalar condensates.
According to our choice $\bar\pi_i= \delta_{i1} \bar \Delta$, we define the
charged pion condensate $\Pi = \langle \bar \psi i \gamma_5 \tau_1 \psi
\rangle$. The analytical expression for this condensate can be obtained by
taking the derivative of the thermodynamic potential with respect to an
auxiliary parameter added to $\rho(\bar p)$ in Eq.~(\ref{actionMF}), and
then set to zero after the calculation~\cite{Carlomagno:2021gcy}.

To study the phase transitions, we also introduce the susceptibilities
associated to the above defined order parameters~\cite{Lu:2019diy}. They
are given by
\begin{equation}
\chi_{\rm ch} \ = \ - \frac{\partial\Sigma}{\partial m_c} \ , \qquad
\chi_{\Pi} \ = \ \frac{\partial\Pi}{\partial m_c} \ .
\label{chis}
\end{equation}

Finally, as mentioned in the Introduction, it is interesting to consider the
speed of sound $c_s$. At zero temperature, one has
\begin{equation}
c_s^2 \ = -\,\frac{\partial \Omega^{\rm MFA,reg}}{\partial \varepsilon}\ ,
\end{equation}
where the energy density $\varepsilon$ is given by
\begin{equation}
\varepsilon \ = \ \Omega^{\rm MFA,reg} + n_I\, \mu_I + n_B\, \mu_B\ ,
\end{equation}
with $n_I = -\,\partial \Omega^{\rm MFA,reg}/\partial \mu_I$,
$n_B = -\, \partial \Omega^{\rm MFA,reg}/\partial \mu_B$.

\section{Numerical Results}
\label{results}

To fully define our model it is necessary to specify the form factor
entering the nonlocal fermion current given by Eq.~(\ref{cuOGE}). In this
work we consider an exponential momentum dependence for the form factor (in
momentum space),
\begin{equation}
g(p) \ = \ \exp (-p^2 / \Lambda^2)\ .
\label{ff}
\end{equation}
This form, which is widely used, guarantees a fast ultraviolet convergence
of quark loop integrals. Notice that the energy scale $\Lambda$, which acts
as an effective momentum cutoff, has to be taken as an additional parameter
of the model. Other functional forms, as e.g.\ Lorentzian form factors with
integer or fractional momentum dependences, have also been considered in the
literature~\cite{Dumm:2010hh,Carlomagno:2018tyk}. In any case, it is seen
that the form factor choice does not have in general major impact in the
qualitative predictions for most relevant thermodynamic
quantities~\cite{Dumm:2021vop}.

Given the form factor shape, the model parameters $m_c$, $G$ and $\Lambda$
can be fixed by requiring that the model can reproduce the phenomenological
values of some selected physical quantities. Here we take as inputs the
empirical values of the pion mass, $m_\pi=138$~MeV, and the pion weak decay
constant, $f_\pi=92.4$~MeV, together with phenomenologically reasonable
values of the quark-antiquark condensates at $\mu=\mu_I=0$, viz.~$\Sigma_u =
\Sigma_d = - (240\ {\rm MeV})^3$. This leads to $m_c = 5.67$~MeV, $\Lambda =
752$~MeV and $G\Lambda^2 = 20.67$~\cite{GomezDumm:2006vz}. In
Ref.~\cite{Carlomagno:2021gcy} this parametrization has been used to study
the features of this type of model for a system at finite temperature and
isospin chemical potential, getting a good agreement with LQCD results for
the $T-\mu_I$ phase diagram. As mentioned in the Introduction, one of the
aims of this work is to confront the results obtained within the nonlocal
model with those obtained in the framework of the standard, local version of
the NJL model. For the latter we use the parametrization $m_c = 5.83$~MeV,
$\Lambda_0 = 588$~MeV and  $G\Lambda_0^2 =4.88$, which leads to $\Sigma_u =
\Sigma_d = - (239\ {\rm MeV})^3$ and quark effective masses $M_u=M_d =
400$~MeV~\cite{Buballa:2003qv}.

\subsection{Results for $\mu_B=0$ and finite $\mu_I$}

We begin by stating the picture obtained from the nonlocal NJL model at
vanishing baryon chemical potential. It is interesting to compare our
results with LQCD calculations, which for $\mu_B = 0$ are free from the sign
problem, even for nonzero values of the isospin chemical potential.

\subsubsection{Order parameters and phase transitions}

\begin{figure}[hbt]
\centering
\includegraphics[width=0.6\textwidth]{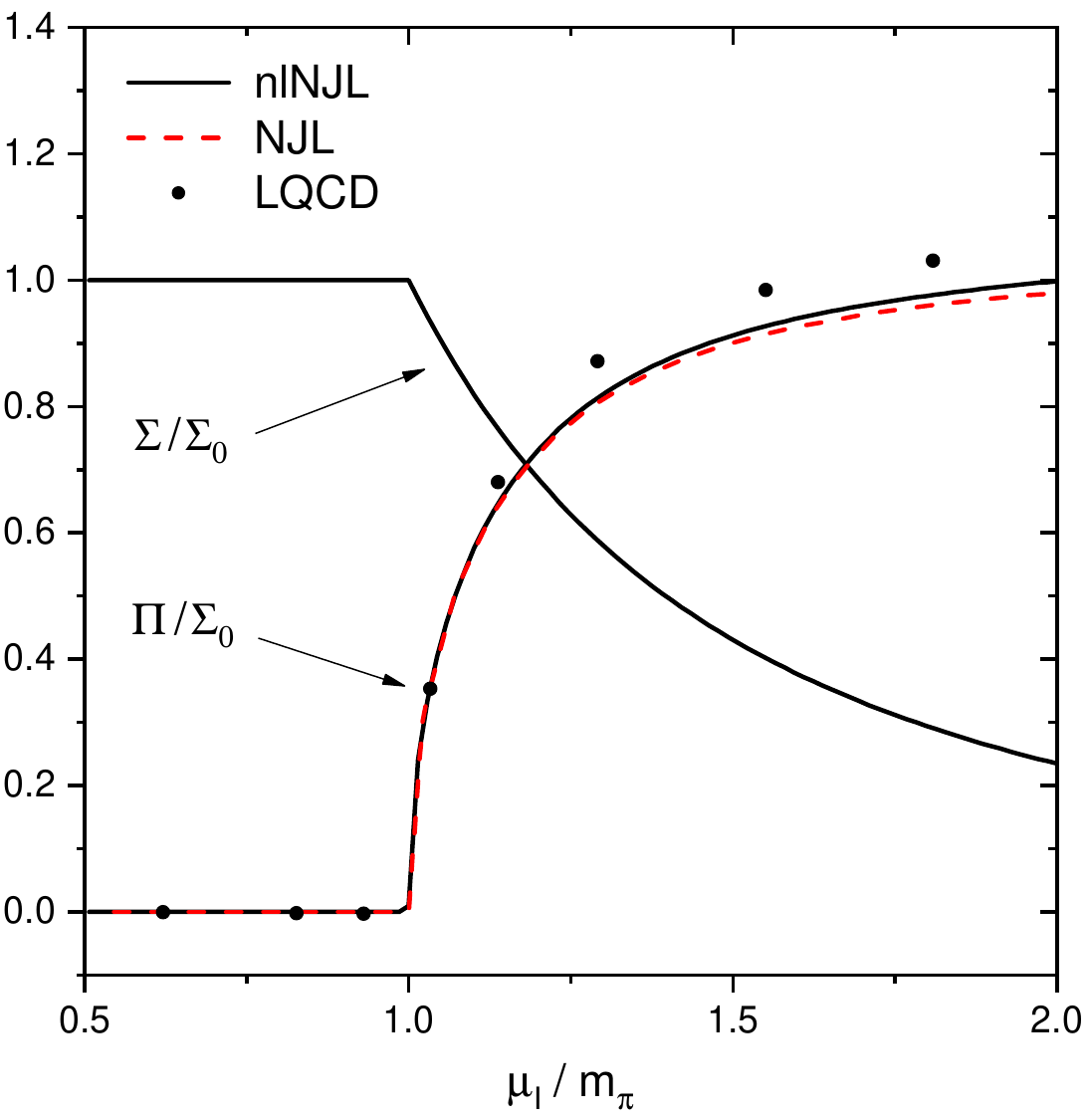}
\caption{\small{Normalized $\Sigma$ and $\Pi$ condensates as
functions of the isospin chemical potential. Lattice results from
Ref.~\cite{Brandt:2018bwq} are included for comparison.}}
\label{fig:mu0}
\end{figure}

In Fig.~\ref{fig:mu0} we show the behavior of the above introduced
$\Sigma$ and $\Pi$ condensates at $\mu_B=0$ and finite $\mu_I$. Although
these results have been previously presented in Ref.~\cite{Carlomagno:2021gcy},
we find it convenient to include a brief review for the sake comparison with
the case of nonzero $\mu_B$.

For $\mu_I < m_\pi$ one finds the usual low energy situation in which chiral
symmetry is spontaneously broken, which is reflected in a large value
$\Sigma = \Sigma_0$ for the quark-antiquark condensate, while the pion
condensate vanishes ($\Pi = \Pi_0 =0$). Then, it can be analytically shown
that at $\mu_I = m_\pi$ the model predicts the onset of a phase in which one
has pion condensation. For $\mu_I > m_\pi$, as seen in Fig.~\ref{fig:mu0},
the chiral condensate decreases monotonically and the charged pion
condensate gets strongly increased. Thus, one has a second order phase
transition in which the isospin symmetry U(1)$_{I_3}$ gets spontaneously
broken, whereas one finds a smooth partial restoration of the U(1)$_{I_3A}$
symmetry when reaching large values of $\mu_I$. It can be seen that the
results from local and nonlocal versions of the NJL model are similar to
each other, and they are found to be in good qualitative agreement with
lattice QCD calculations (also shown in the figure)~\cite{Brandt:2018bwq}.
In addition, as discussed in Ref.~\cite{Carlomagno:2021gcy}, for this range
of values of $\mu_I$ the results are consistent with the relation
\begin{equation}
\left(\frac{\Sigma}{\Sigma_0}\right)^2 +
\left(\frac{\Pi}{\Pi_0}\right)^2\ = \ 1\ ,
\label{chiral}
\end{equation}
which can be obtained from lowest-order chiral perturbation
theory~\cite{Kogut:2001id}.

\subsubsection{Speed of sound}

As mentioned above, the speed of sound $c_s$ has been studied within various
effective models. For the case of systems at nonzero isospin chemical
potential, recent LQCD calculations~\cite{Brandt:2022hwy,Abbott:2023coj}
have found that the curve of $c_s^2$ as a function of $\mu_I$ shows a
maximum for $\mu_I / m_\pi \sim 2$. This maximum is shown to be well above
the limit $c_s^2 = 1/3$, which is obtained for QCD at large temperature on
the basis of gauge/gravity duality for 4D conformal field
theories~\cite{Cherman:2009tw}. It is worth noticing that a similar behavior
has been obtained in the framework of two-color QCD~\cite{Iida:2022hyy,Itou:2022ebw,Kojo:2021hqh} and quarkyonic models
for dense quark matter~\cite{McLerran:2018hbz,Jeong:2019lhv,Kovensky:2020xif,Kojo:2021ugu}.

\begin{figure}[hbt]
\centering
\includegraphics[width=0.7\textwidth]{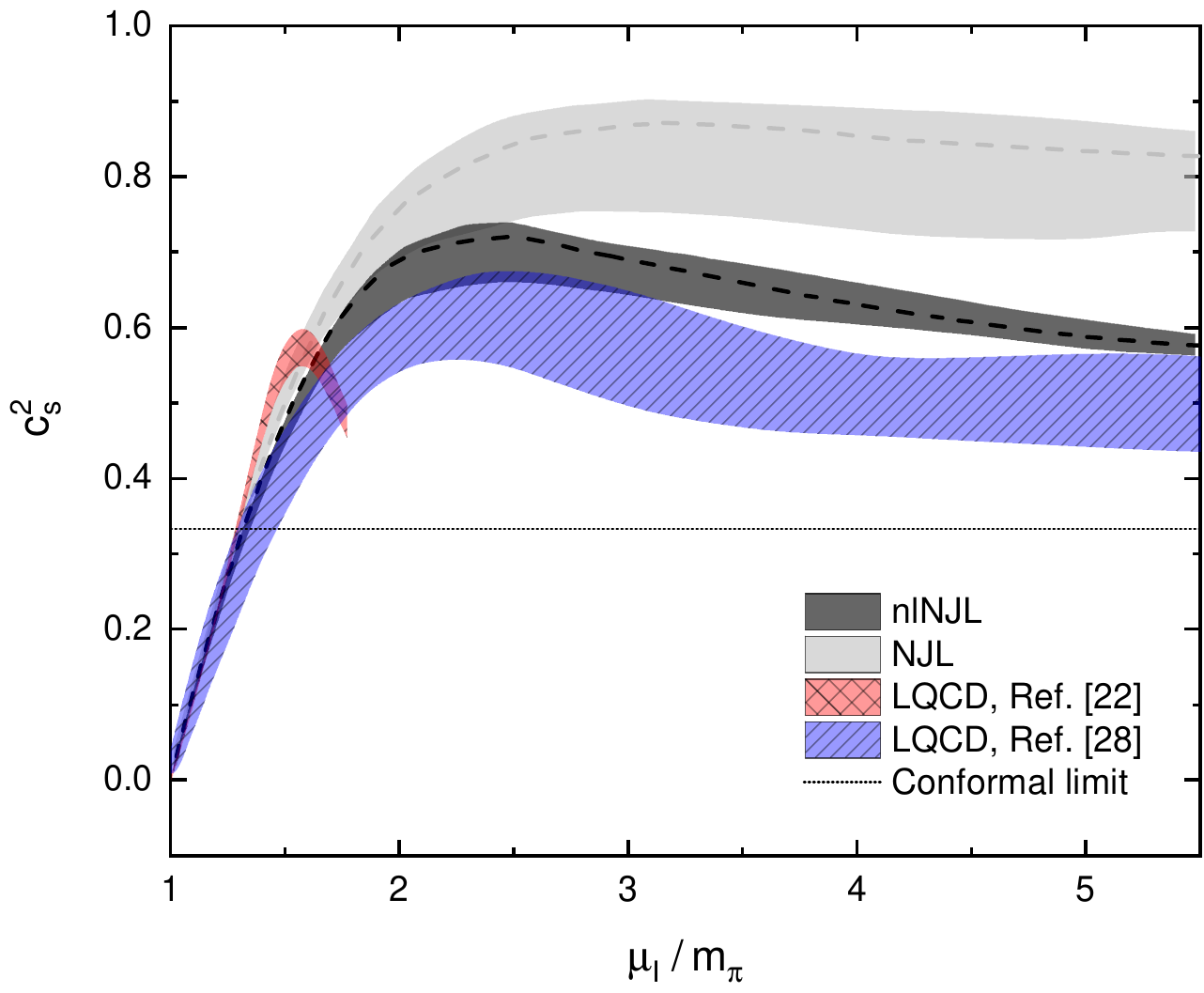}
\caption{\small{Squared speed of sound for $\mu_B=0$. Dark and light gray
bands correspond to the results for local and nonlocal NJL models,
respectively. The dashed lines within the gray bands correspond to
model parameterizations that lead to a quark-antiquark condensate of
$-(240~{\rm MeV})^3$. LQCD data from
Refs.~\cite{Brandt:2022hwy,Abbott:2023coj} are also included for
comparison.}} \label{fig:css}
\end{figure}

Our numerical results are shown in Fig.~\ref{fig:css}, where we also include
for comparison the results arising from the local NJL model and those
obtained from LQCD in Refs.~\cite{Brandt:2022hwy,Abbott:2023coj}. To provide
an estimate of the dependence on model parameterizations, we have considered
the results for the nonlocal NJL model for parameters corresponding to
quark-antiquark condensates lying within the range from $-(260\ {\rm
MeV})^3$ to $-(230\ {\rm MeV})^3$ (dark gray band), and similarly for the
local NJL model, taking a range from $-(250\ {\rm MeV})^3$ to $-(240\ {\rm
MeV})^3$ (light gray band). It can be seen that the results obtained within
the nonlocal model do not show a strong dependence on the parameterization.
Moreover, they reproduce with good qualitative agreement the behavior
observed by the most recent LQCD analysis ---see
Ref.~\cite{Abbott:2023coj}---, where a large range of values of $\mu_I$ is
covered. On the other hand, in agreement with the results in
Ref.~\cite{Ayala:2023mms}, it is seen that for the local NJL model $c_s^2$
does not show a clear peak at intermediate values of $\mu_I$. In fact, in
Ref.~\cite{Ayala:2023mms} it is found that such a peak can be obtained once
the coupling constants are allowed to have an explicit dependence on
$\mu_I$.

\subsection{Phase transitions for finite baryon chemical potential}

\begin{figure}[hbt]
\centering
\hspace{-1.2cm}
\includegraphics[width=0.55\textwidth]{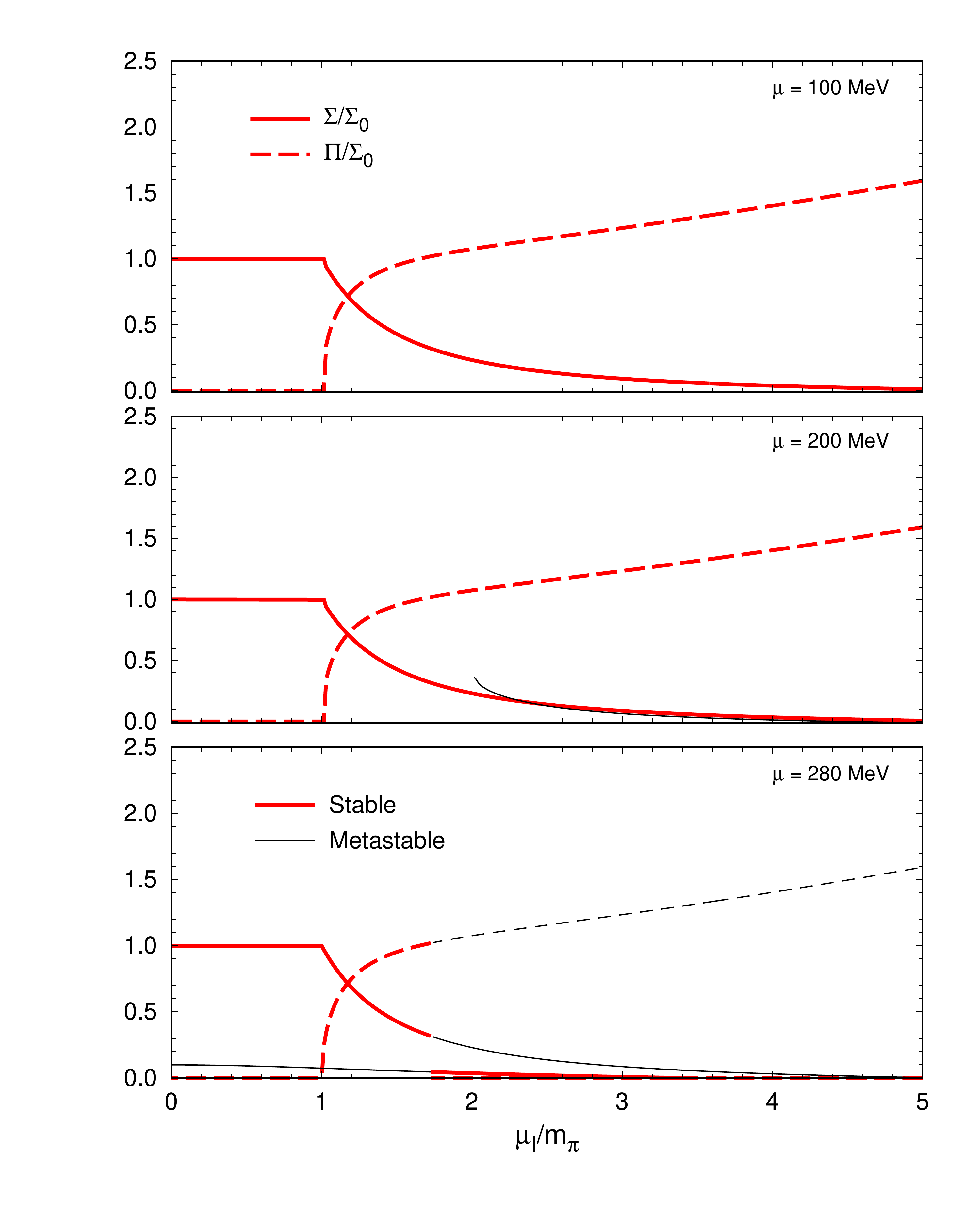}
\hspace{-1.2cm}
\includegraphics[width=0.55\textwidth]{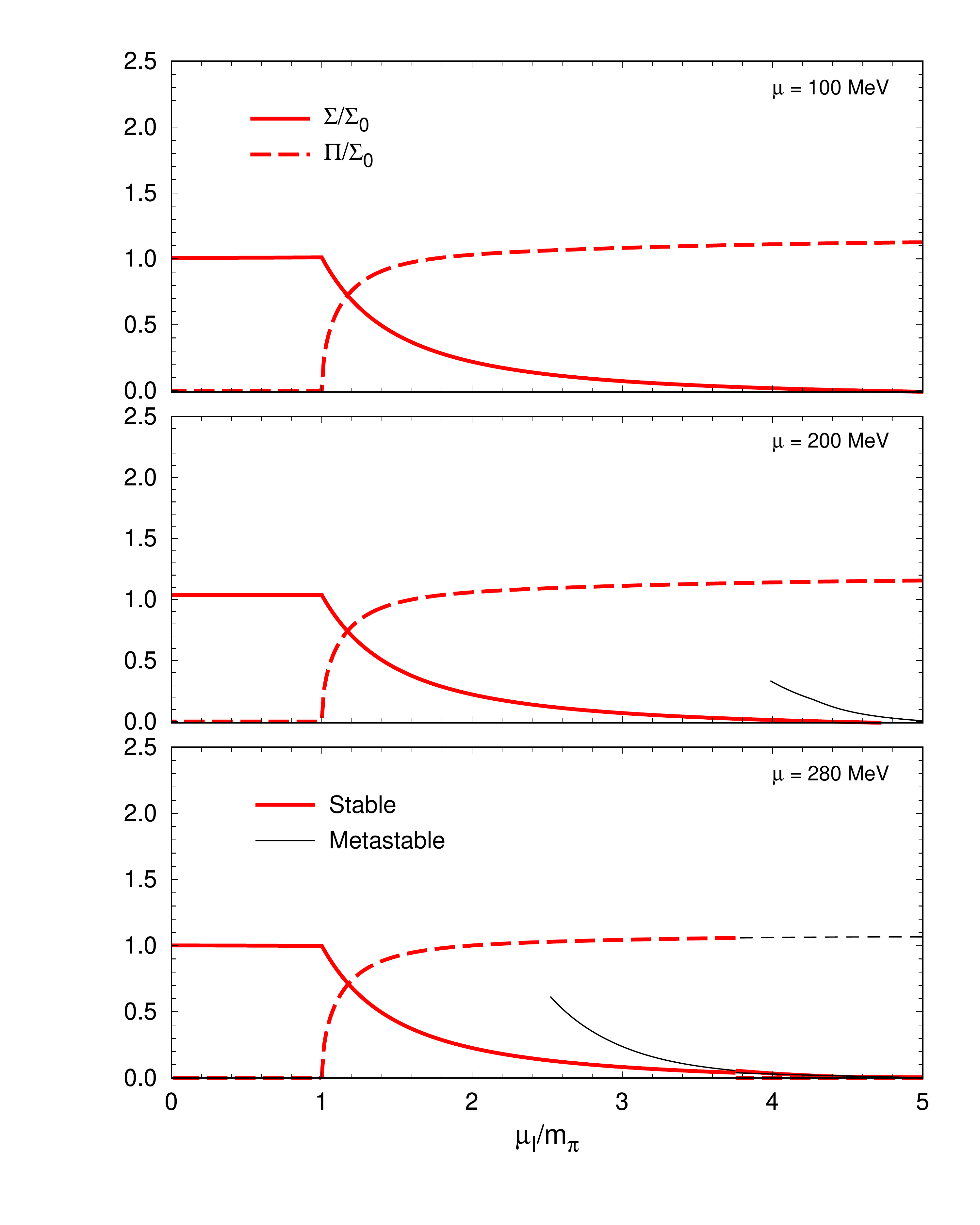}
\caption{\small{Normalized values of the order parameters $\Sigma$ and $\Pi$
as functions of $\mu_I/m_\pi$ for nonlocal (left) and local (right) NJL models.}}
\label{fig:OP1}
\end{figure}

Let us consider a more general situation in which both the quark chemical
potential $\mu = \mu_B/3$ and the isospin chemical potential are nonzero. To
describe the picture obtained in the $(\mu,\mu_I)$ thermodynamic space we
take some representative values of $\mu$ and study how the order parameters
$\Sigma$ and $\Pi$ evolve with $\mu_I$. This is shown in Fig.~\ref{fig:OP1}.
Left and right panels correspond to the results from nonlocal and local NJL
models, respectively.

For low values of $\mu$ the situation is similar to the one described in the
previous section for $\mu=0$. In the left and right upper panels of
Fig.~\ref{fig:OP1} we show the behavior of $\Sigma$ and $\Pi$ for
$\mu=100$~MeV; one finds at $\mu_I/m_\pi = 1$ the onset of a pion
condensation phase (a second order phase transition), while chiral symmetry
gets smoothly restored when $\mu_I$ is increased. Notice that, in the case
of the nonlocal model, for large values of $\mu_I$ there is a significant
deviation from the chiral relation in Eq.~(\ref{chiral}); this deviation is
not observed for the local model (at least, for values of $\mu_I$ up
to $5\,m_\pi$).

\begin{figure}[hbt]
\centering
\includegraphics[width=0.95\textwidth]{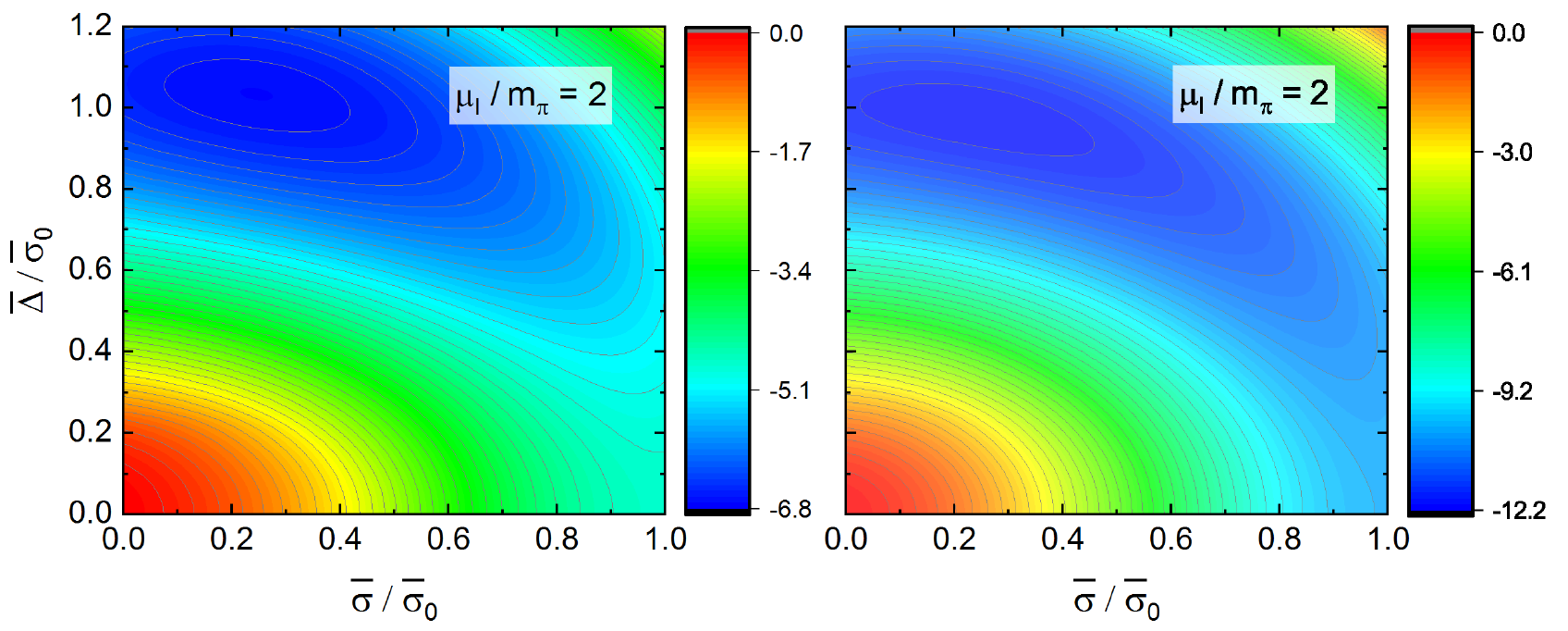}
\caption{\small{Contour plots for $\Omega^{\rm
MFA,reg}(\bar\sigma,\bar\Delta)\times 10^4$ (in GeV$^4$) for
$\mu=100$~MeV as functions of $\bar\sigma/\bar\sigma_0$ and
$\bar\Delta/\bar\sigma_0$. Left and right panels correspond to NJL nonlocal and
local models, respectively.}}
\label{fig:mu100}
\end{figure}

To get a better understanding of the transitions let us also show contour
plots that describe the behavior of the mean field thermodynamic potential
$\Omega^{\rm MFA,reg}$ as a function of the VEVs $\bar\sigma$ and
$\bar\Delta$ for particular values of $\mu_I$. In Fig.~\ref{fig:mu100} we
consider the case $\mu = 100$~MeV, $\mu_I/m_\pi = 2$. Left and right panels
correspond to the nonlocal and local models, respectively. It is seen that,
for both models, at $(\bar\sigma/\bar\sigma_0,\bar\Delta/\bar\sigma_0)
\simeq (0.25,1)$ one finds a solution of the gap equations (\ref{gapeqs})
that minimizes the thermodynamic potential, while a maximum of $\Omega^{\rm
MFA,reg}$ is obtained at $(\bar\sigma/\bar\sigma_0,\bar\Delta/\bar\sigma_0)
= (0,0)$. The situation is found to be qualitatively similar for larger
values of $\mu_I$.

\begin{figure}[hbt]
\centering
\includegraphics[width=0.95\textwidth]{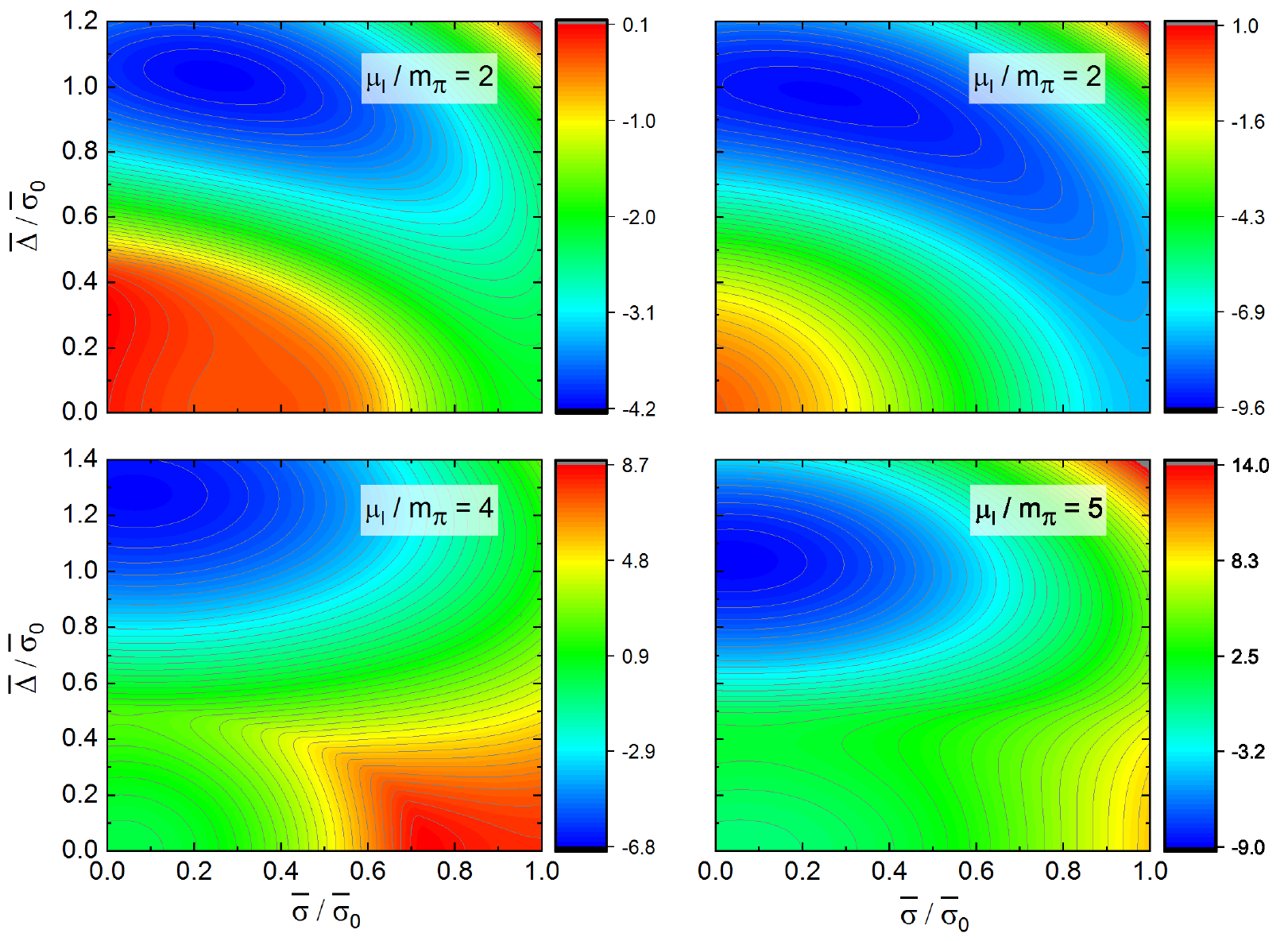}
\caption{\small{Contour plots for $\Omega^{\rm
MFA,reg}(\bar\sigma,\bar\Delta)\times 10^4$ (in GeV$^4$) for
$\mu=200$~MeV as functions of $\bar\sigma/\bar\sigma_0$ and
$\bar\Delta/\bar\sigma_0$. Left and right panels correspond to NJL nonlocal and
local models, respectively.}}
\label{fig:mu200}
\end{figure}

Next, in Fig.~\ref{fig:mu200} we show contour plots for $\mu=200$~MeV,
considering the values $\mu_I/m_\pi = 2$ and $\mu_I/m_\pi = 4 (5)$ for the
nonlocal (local) model. Comparing top to bottom panels, it is seen that the
maximum located at the axis $\bar\Delta =0$ moves to large values of
$\bar\sigma$, and a second local minimum arises close to the origin. In
addition, a saddle point is shown to arise between both local minima. The
second minimum represents a metastable point for which there is no pion
condensation ($\Pi=0$); it corresponds to the dashed lines in the central
panels of Fig.~\ref{fig:OP1}. The onset of this metastable solution occurs
at some critical isospin chemical potential that we denote by $\mu_I^{\rm
(sp)}$; for the chosen parameterizations, is seen that $\mu_I^{\rm
(sp)}(\mu=200\ {\rm MeV})\simeq 2\, m_\pi$ and $4\, m_\pi$ for the nonlocal
and local models, respectively. We notice that a similar picture has been
obtained in Ref.~\cite{Liu:2021gsi} for the case of a three-flavor NJL
model. However, in that article the saddle point is interpreted as maximum,
since only the dependence of the thermodynamical potential with $\bar\Delta$
is analyzed.

\begin{figure}[hbt]
\centering
\includegraphics[width=0.95\textwidth]{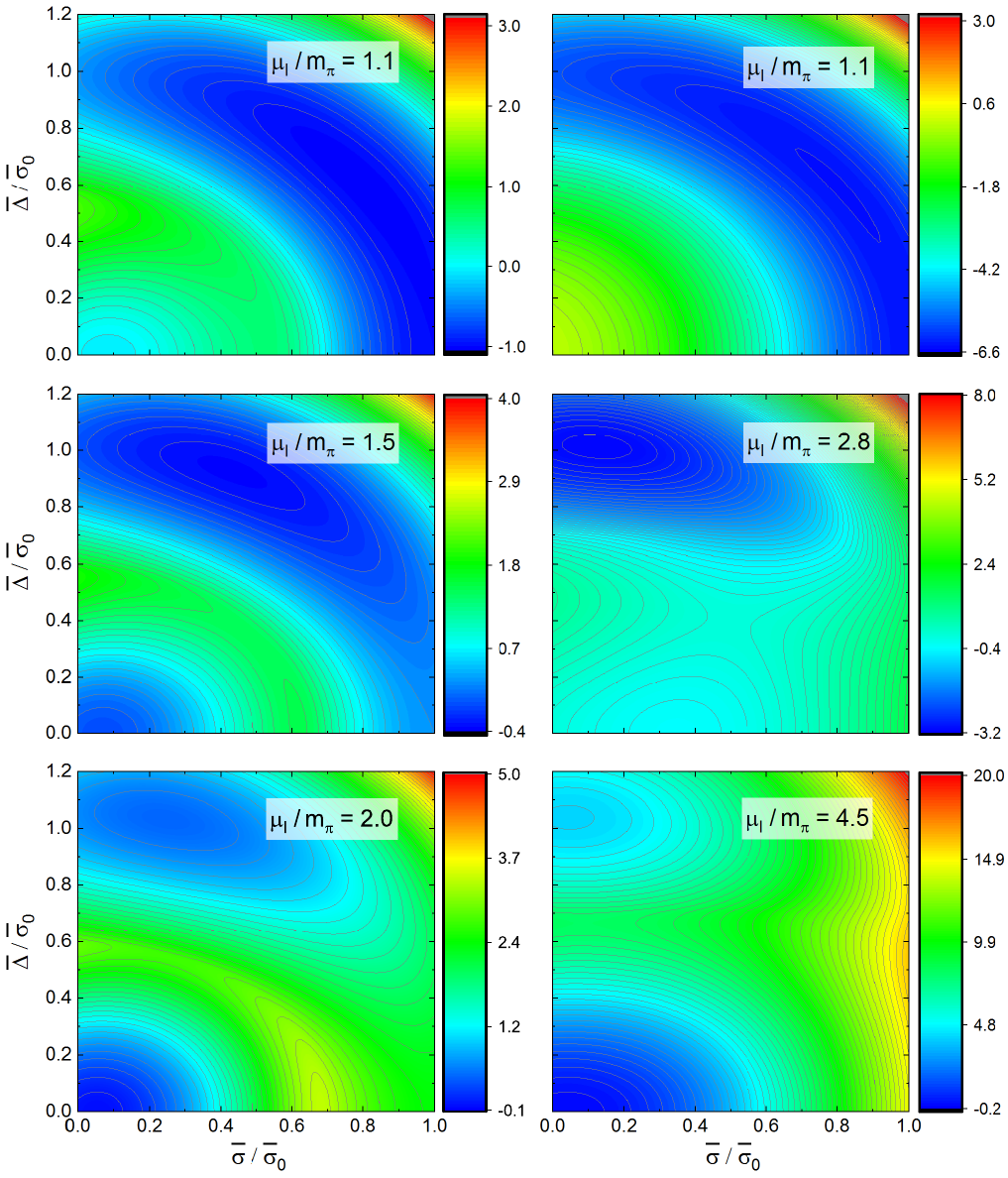}
\caption{\small{Contour plots for $\Omega^{\rm
MFA,reg}(\bar\sigma,\bar\Delta)\times 10^4$ (in GeV$^4$) for
$\mu=280$~MeV as functions of $\bar\sigma/\bar\sigma_0$ and
$\bar\Delta/\bar\sigma_0$. Left and right panels correspond to NJL nonlocal and
local models, respectively.}}
\label{fig:mu280}
\end{figure}

Finally, in Fig.~\ref{fig:mu280} we show how this picture evolves when we go
forward to larger values of the chemical potential $\mu$. To illustrate the
situation we include some contour plots in which we take $\mu = 280$~MeV and
some representative values of $\mu_I/m_\pi$. As in the previous cases, left
(right) panels correspond to the nonlocal (local) NJL model. For values or
$\mu_I$ just above $m_\pi$, in the case of the nonlocal model the metastable
solution already exists (in fact, there is no critical value $\mu_I^{\rm
(sp)}$ for values of $\mu$ larger than about 270~MeV), while for the local
model the situation is similar as the one shown in the upper panel of
Fig.~\ref{fig:mu200}. For larger values of $\mu_I/m_\pi$ (central panels of
Fig.~\ref{fig:mu280}), in the case of the nonlocal model the second local
minimum becomes deeper, while this second solution also arises for the local
model. Then, if $\mu_I/m_\pi$ is further increased, at some critical value
$\mu_{I,c}$ a first order phase transition occurs: as illustrated in the
lower panels of Fig.~\ref{fig:mu280}, the minima for which one has
$\bar\Delta = 0$ are the ones that become energetically favored; thus, the
system jumps into a phase in which there is no pion condensation and the
$U(1)_{I_3}$ symmetry gets restored. The behavior of the order parameters
for $\mu=280$~MeV is shown in the lower panels of Fig.~\ref{fig:OP1}. In the
case of the nonlocal model, it is seen that at the first order transition
the value of the quark-antiquark condensate $\Sigma$ shows also a jump that
implies an approximate restoration of chiral symmetry (in the case of the
local model, the value of $\Sigma$ is already very low when the transition
is reached).

\begin{figure}[hbt]
\centering
\hspace*{0.cm}
\includegraphics[width=0.65\textwidth]{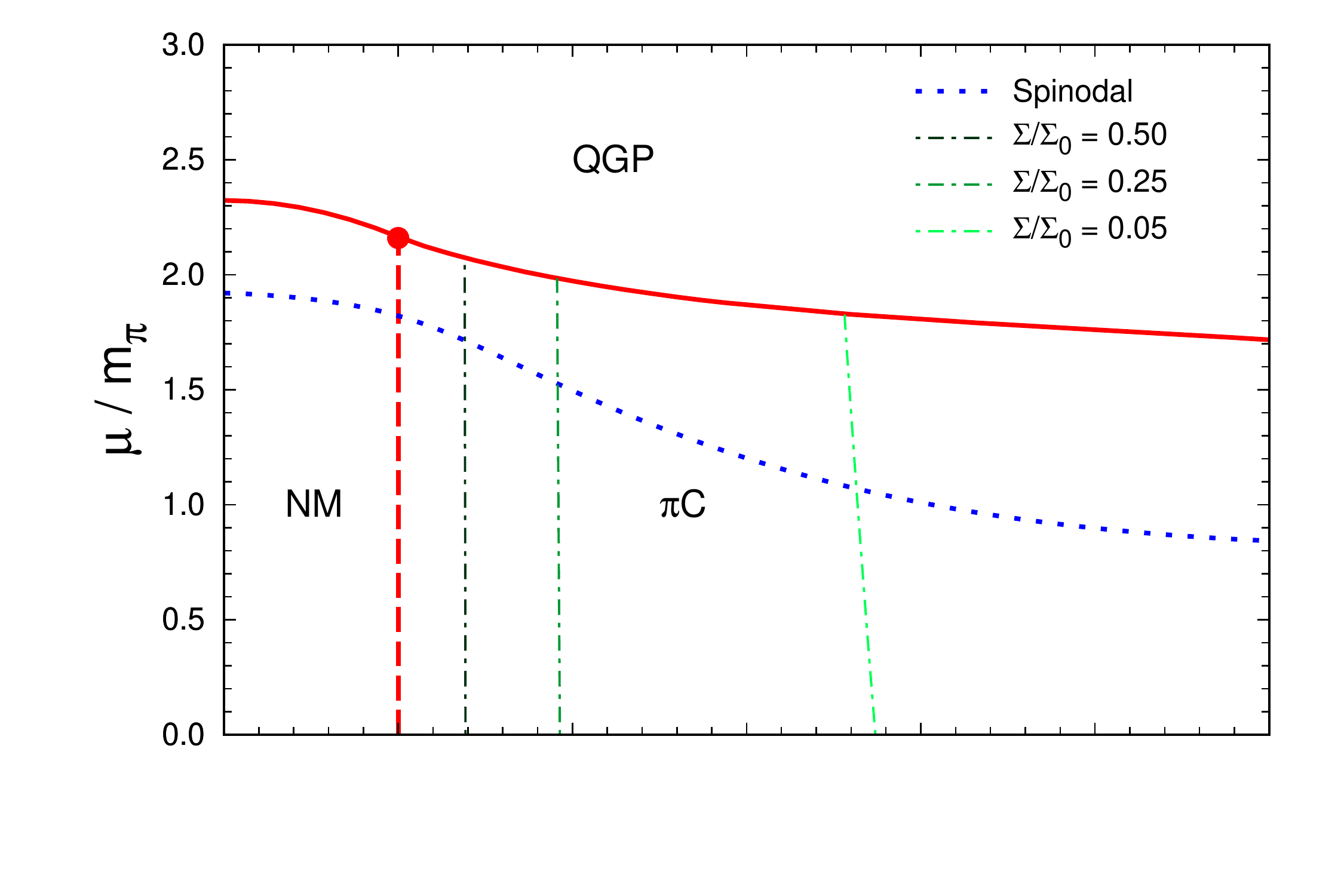}
\vspace*{-1.2cm}\\
\includegraphics[width=0.65\textwidth]{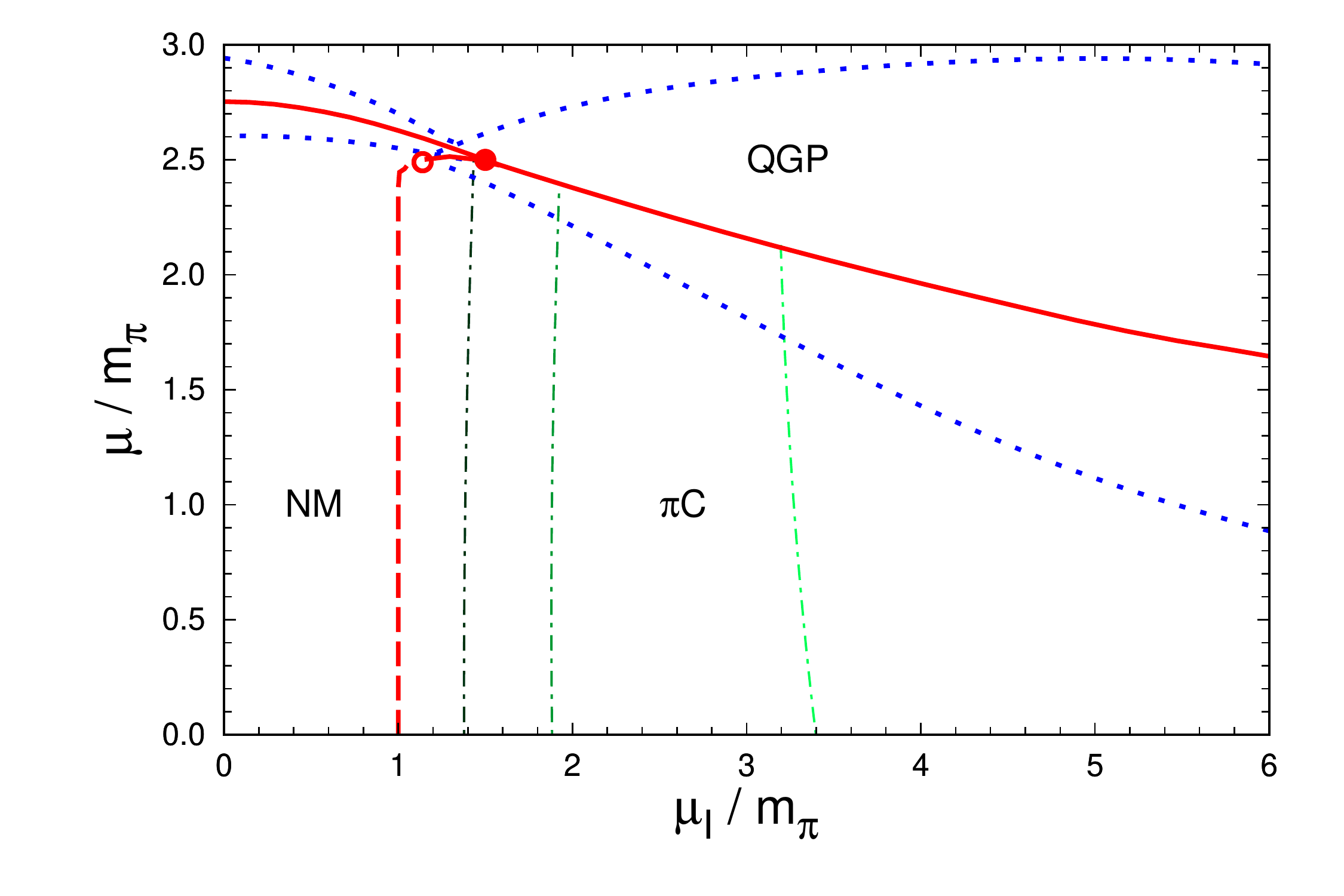}
\caption{\small{QCD phase diagram in the $\mu_I-\mu$ plane at zero
temperature for the nonlocal (left) and local NJL models. Solid (dashed)
lines correspond to first (second) order phase transitions. NM, $\pi$C and
QGP stand for normal hadronic matter, pion condensation and quark-gluon
plasma, respectively.}}
\label{fig:PD}
\end{figure}

For even larger values of $\mu$ the region in which there is a stable
nonvanishing pion condensate gets subsequently reduced, until one reaches a
triple point in which three phases coexist. The full phase diagrams in the
$\mu-\mu_I$ plane for both nonlocal (upper panel) and local (lower panel)
models are shown in Fig.~\ref{fig:PD}. Solid and dashed lines denote first
and second order phase transitions, respectively, while dotted lines denote
the spinodals~\cite{GomezDumm:2004sr} ---boundaries of the region in which
energetically unfavored solutions exist as metastable states.

The phase diagrams show a region of normal matter (NM) in which one has
$\Pi=0$ and the chiral symmetry is spontaneously broken, a quark-gluon
plasma phase in which $\Pi=0$ and the chiral symmetry is approximately
restored, and a region in which one finds pion condensation ($\pi$C),
characterized by a nonvanishing value of $\Pi$. In this last region the
U(1)$_{I_3A}$ symmetry is progressively restored when $\mu_I$ is increased.
As discussed above, for low values of $\mu$ the onset of the $\pi$C phase is
found to occur at $\mu_I = m_\pi$ as a second order phase transition (dashed
lines in the figure). The progressive decrease of the chiral order parameter
$\Sigma$ with $\mu_I$ (with the consequent partial restoration of the
U(1)$_{I_3A}$ symmetry) is illustrated through the dash-dotted lines, which
correspond to constant ratios $\Sigma/\Sigma_0 =0.5$, 0.25 and 0.05. In this
way, for values of $\mu_I$ larger enough than $m_\pi$, the first order
transition indicated by the solid red line occurs between a phase of
(almost) massless asymptotically free quarks and a phase of (almost) pure
pion condensation.

The filled dots in the figure indicate triple points where two transition
lines meet. Notice that in the case of the nonlocal model the second order
and first order transition lines meet exactly at $\mu_I=m_\pi$. In contrast,
for the local model, the second order transition line has a critical
endpoint (open dot in the lower panel of Fig.~\ref{fig:PD}) where it becomes
of first order, and then, for $\mu_I > m_\pi$, it smoothly merges the first
order chiral restoration line. Thus, in a narrow region $1.2\simeq
\mu_I/m_\pi\simeq 1.6$ one can find two first order phase transitions when
$\mu$ is increased: at $\mu\sim 350$~MeV one has a transition from the
$\pi$C phase to the NM phase, followed by a transition to the QGP phase. In
fact, these features of the phase diagram for the local NJL model are in
agreement with the results obtained in
Refs.~\cite{Andersen:2007qv,Liu:2021gsi} for two- and three-flavor (local)
NJL models.

In Fig.~\ref{fig:PD} we also show the spinodals, represented by blue dotted
lines. As stated, these lines indicate the critical isospin chemical
potentials $\mu_I^{\rm (sp)}$ at which metastable solutions are found to
appear. For the local NJL model we find spinodals at both sides of the first
order transition line ---delimiting a band in the phase diagram where
metastable solutions exist---, while in the case of the nonlocal model no
upper spinodal is found. In fact, for large values of $\mu$ one can always
found a metastable solution in which quarks have a relatively large
effective dynamical mass. This is a well-known feature of nonlocal NJL
models; it is related to the fact that ---depending on the model
parameterization--- the quark propagators may not have purely real poles in
Minkowski space~\cite{GomezDumm:2001fz}.

If one goes further to larger values of $\mu_I$, the nlNJL and NJL
approaches show significant differences. For low values of $\mu$, the order
parameter $\Pi$ increases monotonically for the nlNJL model, while for the
local model (as shown in Ref.~\cite{Liu:2023uxm}) it starts to decrease and
goes to zero at some point beyond $\mu_I\sim 10\,m_\pi$. We understand that
these different behaviors are artifacts that arise from the regularization
of the (nonrenormalizable) models, which become hardly trustable in that
limit. Therefore we present our results for more conservative values of
$\mu_I$, up to a few $m_\pi$.

\section{Summary and conclusions}
\label{summary}

The phase diagram of strongly interacting matter has been examined in
two-flavor NJL-type models, considering zero temperature and nonzero
baryon and isospin chemical potentials. Specifically, we have investigated
the transitions related to the order parameters $\Sigma$ and $\Pi$, which
characterize the spontaneous breakdowns of chiral and isospin symmetries,
in models with local and non-local four-quark interactions. We have also
studied the behavior of the speed of sound as a function of the isospin
chemical potential.

Considering a $\mu-\mu_I$ phase diagram, for baryon chemical potentials
lower than about $280$~MeV and $\mu_I < m_\pi$ one finds a region of normal
hadronic matter, in which chiral symmetry is spontaneously broken. Then, at
$\mu_I = m_\pi$ there is a second order phase transition into a region in
which one has a nonzero charged pion condensate. By increasing $\mu_I$ the
chiral condensate $\Sigma$ gets progressively reduced, implying a smooth
restoration of the U(1)$_{I_3A}$ symmetry. On the other hand, by increasing
$\mu$ one arrives at a first order transition to a quark-gluon plasma phase
in which there is no pion condensation and chiral symmetry is approximately
restored. In the nonlocal NJL approach it is seen that the first and second
order transition meet at a triple point located at $\mu_I=m_\pi$, $\mu\simeq
300$~MeV. In the case of the local model, in agreement with previous works,
we find that the second order transition line has a critical endpoint where
it becomes of first order, and then, for $\mu_I > m_\pi$, it smoothly merges
the first order chiral restoration line. We have also studied metastable
phases, identifying saddle points for the thermodynamic potential as a
function of the order parameters; in general, it is seen that metastable
phases cover a larger region in the case of the nonlocal model. Concerning
the speed of sound $c_s$, we find that for the nonlocal model the behavior
of $c_s^2$ with $\mu_I$ for vanishing baryon chemical potential shows a
maximum at $\mu_I\sim 2\,m_\pi$, improving the qualitative agreement with
lattice QCD calculations in comparison with the results obtained for the
local NJL approach. It would be interesting to extend these studies to
systems at finite temperature, with the aim of determining the behavior of
the triple point and the phase transition lines (the case $\mu=0$ has been
already considered in Ref.~\cite{Carlomagno:2021gcy}). In addition, it would
be worthwhile to consider the case of neutral matter conditions, to be
applied to the composition of compact objects like neutron or pion stars.

\section*{Acknowledgements}

This work has been supported in part by Consejo Nacional de Investigaciones
Cient\'ificas y T\'ecnicas and Agencia Nacional de Promoci\'on Cient\'ifica
y Tecnol\'ogica (Argentina), under Grants No.~PIP2022-GI-11220210100150CO,
No.~PICT-2019-00792 and No.~PICT20-01847, and by the National University of
La Plata (Argentina), Project No.~X960.



\end{document}